\RequirePackage{lineno}
\documentclass[twocolumn,aps,floatfix,prc,twocolumn,showpacs,showkeys,amsmath,amssymb,superscriptaddress]{revtex4-1}

\usepackage{soul}
\usepackage{graphicx}	\usepackage{xspace}     \usepackage{amsmath}
\usepackage{hyperref}\usepackage{color}\usepackage{multirow}\usepackage{booktabs,array}

\makeatletter{}\newcommand{\jth}{$j$th\xspace}
\newcommand{\jhcs}{jet-hadron correlations\xspace}
\newcommand{\dhcs}{di-hadron correlations\xspace}
\newcommand{\dhjhcs}{di-hadron and jet-hadron correlations\xspace}
\newcommand{\Dhjhcs}{Di-hadron and jet-hadron correlations\xspace}

\newcommand{\pT}{$p_{\rm T}$\xspace}

\newcommand{\dphi}{$\Delta\phi$\xspace}

\newcommand{\GeV}{GeV/$c$\xspace}

\newcommand{\vn}{$v_{n}$\xspace}
\newcommand{\vnt}{$v_{n}^{t}$\xspace}

\newcommand{\vnum}[1]{$v_{#1}$\xspace}

\newcommand{\vnumeff}[1]{$\tilde{v}_{#1}$\xspace}

\newcommand{\eref}[1]{equation~\ref{#1}}

\newcommand{\Fref}[1]{Figure~\ref{#1}}

\newcommand{\pp}{$p$+$p$\xspace}

\begin{document}

\title{ Event plane dependence of the flow modulated background in di-hadron and jet-hadron correlations in heavy ion collisions}

\author{Christine Nattrass} 
\affiliation{University of Tennessee, Knoxville, TN, USA-37996.}
\author{Takahito Todoroki}
\affiliation{Physics Department, Brookhaven National Laboratory, Upton, New York 11973-5000, USA.}
\affiliation{RIKEN BNL Research Center, Brookhaven National Laboratory, Upton, New York 11973-5000, USA}
\date{\today}

\begin{abstract} 
\Dhjhcs are commonly used in relativistic heavy ion collisions to study the soft component of jets in a quark gluon plasma.  There is a large correlated background which is described by the Fourier decomposition of the azimuthal anisotropy where \vn is the $n$th order coefficient.  The path length dependence of partonic energy loss can be studied by varying the angle of the high momentum trigger particle or jet relative to a reconstructed event plane.  This modifies the shape of the background correlated with that event plane.  The original derivation of the shape of this background only considered correlations relative to the second order event plane, which is correlated to the initial participant plane.  We derive the shape of this background for an event plane at an arbitrary order.  There is a phase shift in the case of jets restricted to asymmetric regions relative to the event plane. 
For realistic correlations between event planes, the correlation between the second and fourth order event planes leads to a much smaller effect than the finite event plane resolution at each order.  Finally, we assess the status of the rapidity even \vnum{1} term due to flow, which has been measured to be comparable to \vnum{2} and \vnum{3} terms.

\end{abstract}

  \maketitle

 \section{Introduction}\label{Sec:Introduction}
\makeatletter{}A hot and dense medium called a Quark Gluon Plasma (QGP) is formed in high energy heavy ion collisions~\cite{Adcox:2004mh,Adams:2005dq,Back:2004je,Arsene:2004fa}.  Two primary signatures of the QGP are hydrodynamical flow and jet quenching.  Hydrodynamical flow leads to an azimuthally asymmetric distribution of final state hadrons due to asymmetric pressure gradients in the medium~\cite{Adam:2016izf,Chatrchyan:2013kba,Aad:2014vba,Adler:2003kt,Alver:2006wh,Adler:2001nb}.  This is quantified by flow harmonics 
$v_n = \langle \cos(n(\phi-\psi_n)) \rangle$, where $n$ is an integer, $\phi$ is the azimuthal angle of the particle, and $\psi_n$ is the azimuthal angle of 
the $n$ th order event plane.  Partonic energy loss in the medium is shown by the suppression of particle production relative to that in \pp collisions. This suppression also leads to azimuthal asymmetries in final state hadrons because the geometry of the colliding nuclei produces
an asymmetry in the path lengths traversed by hard partons~\cite{Connors:2017ptx}.

At low transverse momenta \pT ($p_T \lesssim 1$ \GeV), particle production is dominated by soft processes, with correlations between the event plane due to hydrodynamical flow. At high transverse momenta ($p_T \gtrsim 5$ \GeV) particle production is dominated by jets, leading to correlations with the event plane due to the path length dependent energy loss.  Hard and soft processes can be studied separately in these regimes, however a complete understanding of jet quenching requires disentangling effects from jet production and hydrodynamical flow at intermediate and low momenta because these momentum ranges are where the soft products from processes such as gluon bremsstrahlung appear.

Di-hadron~\cite{Adler:2002tq,Adler:2005ad,Abelev:2009af,Aamodt:2011vg,Alver:2009id} and \jhcs~\cite{Khachatryan:2016erx,Adamczyk:2013jei} are often used in order to study the soft components of jets in heavy ion collisions, studies which require precision background subtraction due to the large combinatorial background.  The background has usually been determined using the Zero-Yield-At-Minimum method~\cite{Adams:2005ph} combined with an assumption that the \vn contributions in correlations are the same as those measured independently.  The shape of this background when the trigger particle or jet is fixed relative to the second-order event plane was derived in~\cite{Bielcikova:2003ku} and was used for studies of the path length dependence of partonic energy loss~\cite{Agakishiev:2014ada,Adare:2010mq}.  The change in this shape with the angle of the trigger particle relative to the event plane can be used to fit both the background level and shape from the correlations themselves~\cite{Sharma:2015qra}; this method was applied to data in~\cite{Nattrass:2016cln}.

There have been several developments since the derivation in~\cite{Bielcikova:2003ku} which have advanced our understanding of correlations due to flow.  While the reaction plane is well-defined as the plane connecting the beam axis and containing the center of both incoming nuclei, we now know that we experimentally measure event planes, the axes of symmetry of the final state particles emitted from the nucleus collisions~\cite{Alver:2010gr,Sorensen:2010zq}.  The event planes of different orders are only partially correlated with each other~\cite{Aad:2014fla}.

We revisit the form of two particle correlations due to flow derived in~\cite{Bielcikova:2003ku} for studies where a trigger particle is fixed relative to an event plane.  We extend the derivation in~\cite{Bielcikova:2003ku} to an arbitrary event plane and consider the impact of correlation between event planes of different orders.  There is a phase shift when asymmetric regions relative to the event plane are studied, not generally of interest for studies of hydrodynamical flow but of potential interest for studies of jets.  We assess the impact of these equations on studies of \dhjhcs and provide some guidance for future studies.

\makeatletter{}\section{Correlations due to flow}\label{Sec:Correlations}
In~\cite{Bielcikova:2003ku}, it was assumed that the density of overlapping regions was determined by the average distributions, neglecting fluctuations in the positions of the nucleons.  We now know that the experimentally reconstructed event plane originates from the distribution of nucleons which participate in the collision, called the participant plane.  The second order event plane corresponds to the reaction plane if nucleons were in their average positions.  The derivations in~\cite{Bielcikova:2003ku} then are for the second order plane.  
\noindent The different orders of event planes are only partially correlated with each other~\cite{Aad:2014fla}.  Since the even order event planes are dominantly from the average nucleon positions, these event planes are strongly correlated with each other, while the odd participant planes are nearly uncorrelated with other orders.

In a typical di-hadron or jet-hadron correlation measurement, a high momentum trigger particle or reconstructed jet is used to define the coordinate system and the distribution of associated particles relative to that trigger particle is measured. The shape of the correlations when the trigger is restricted in angle relative to the event plane can be derived from the azimuthal distribution of single particles or jets
\begin{equation}
 \frac{dN}{d(\phi-\psi_j)} = \frac{N}{2\pi} \Big (1+2 \sum_{n=1}^{\infty} v_{n} \cos(n(\phi-\psi_n)) \Big ) 
\end{equation}
\noindent by taking the product of the distribution of triggers and associated particles.
Note that the \vn can arise due to either flow or any other process, including jet quenching, which leads to a correlation with the event plane -- the shape only depends on correlations with the event plane, not the physical origin of those correlations.  
The derivation of the background level and azimuthal distribution of particles relative to each other 
$\Delta\phi = \phi^a - \phi^t$ when the trigger azimuthal angle relative to the $j$th order event plane $\phi_s=\phi^{t} - \psi_j$ is restricted to $\phi_s-c < \phi_s < \phi_s+c$ can be found in the appendix. The azimuthal distribution of the background is given by
\begin{widetext}
\begin{equation}
 B(\Delta\phi) = \tilde{B}\Bigg(1+
 2\sum_{n=1}^{\infty} v_{n}^{a} \big( \tilde{v}_{n}^{t} \cos(n \Delta\phi) + \tilde{w}_{n}^{t} \sin(n \Delta\phi)\big)
\Bigg).\label{Eq:BFEP1}
\end{equation}
\noindent where
\begin{align}
 \tilde{B} &= \frac{N^t N^a j c}{2\pi^2} \Big(1
 +  2 \sum_{k=1}^{\infty}  \frac{v_{jk}^{t}  }{jkc} \sin(jkc) R_{jk,j} C_{jk,0,j} \cos(jk\phi_s)
 \Big),\nonumber \\ 
 \tilde{v}_{n}^{t} &= 
 \frac{
 v_{n} + \frac{\delta_{n,mult\ j} }{nc} \sin(nc) R_{n,j} C_{n,0,j}   \cos(n\phi_s)+ \sum_{k=1}^{\infty} (v_{jk+n}^{t}C_{|jk+n|,n,j}+v_{|jk-n|}^{t}C_{|jk-n|,n,j}) \frac{\sin(jkc)\cos(jk\phi_s) R_{jk,j}}{jkc}}{
 1 +  2 \sum_{k=1}^{\infty}  \frac{v_{jk}^{t}  }{jkc} \sin(nc) R_{jk,j} C_{jk,0,j} \cos(jk\phi_s)
 }\nonumber \\ \label{Eq:BFE}
 \tilde{w}_{n}^{t} &= \frac{
 \frac{\delta_{n,mult\ j} }{nc} \sin(nc) R_{n,j} C_{n,0,j}   \sin(n\phi_s)+ \sum_{k=1}^{\infty}  (v_{jk+n}^{t}C_{|jk+n|,n,j}+v_{|jk-n|}^{t}C_{|jk-n|,n,j}) \frac{\sin(jkc)\sin(jk\phi_s) R_{jk,j}}{jkc}}{
 1 +  2 \sum_{k=1}^{\infty}  \frac{v_{jk}^{t}  }{jkc} \sin(nc) R_{jk,j} C_{jk,0,j} \cos(jk\phi_s)
 }\\ \nonumber
 R_{n,j} &= \langle \cos(n \Delta\psi_{j}^{reco}) \rangle  = \langle \cos(n (\psi_{j}^{reco} - \psi_{j}^{true})) \rangle\\ \nonumber
 C_{n,m,j} &= \langle \cos(n\psi_{n}+m\psi_{m}-(n+m)\psi_{j}) \rangle \nonumber
\end{align}
where $N^t$ is the number of triggers, $N^a$ is the number of associated particles, $v_{n}^{a}$ are the \vn of the associated particles, and $v_{n}^{t}$ are the \vn of the triggers.  
\end{widetext}

\begin{figure}
\begin{center}
\rotatebox{0}{\resizebox{8cm}{!}{
	\includegraphics{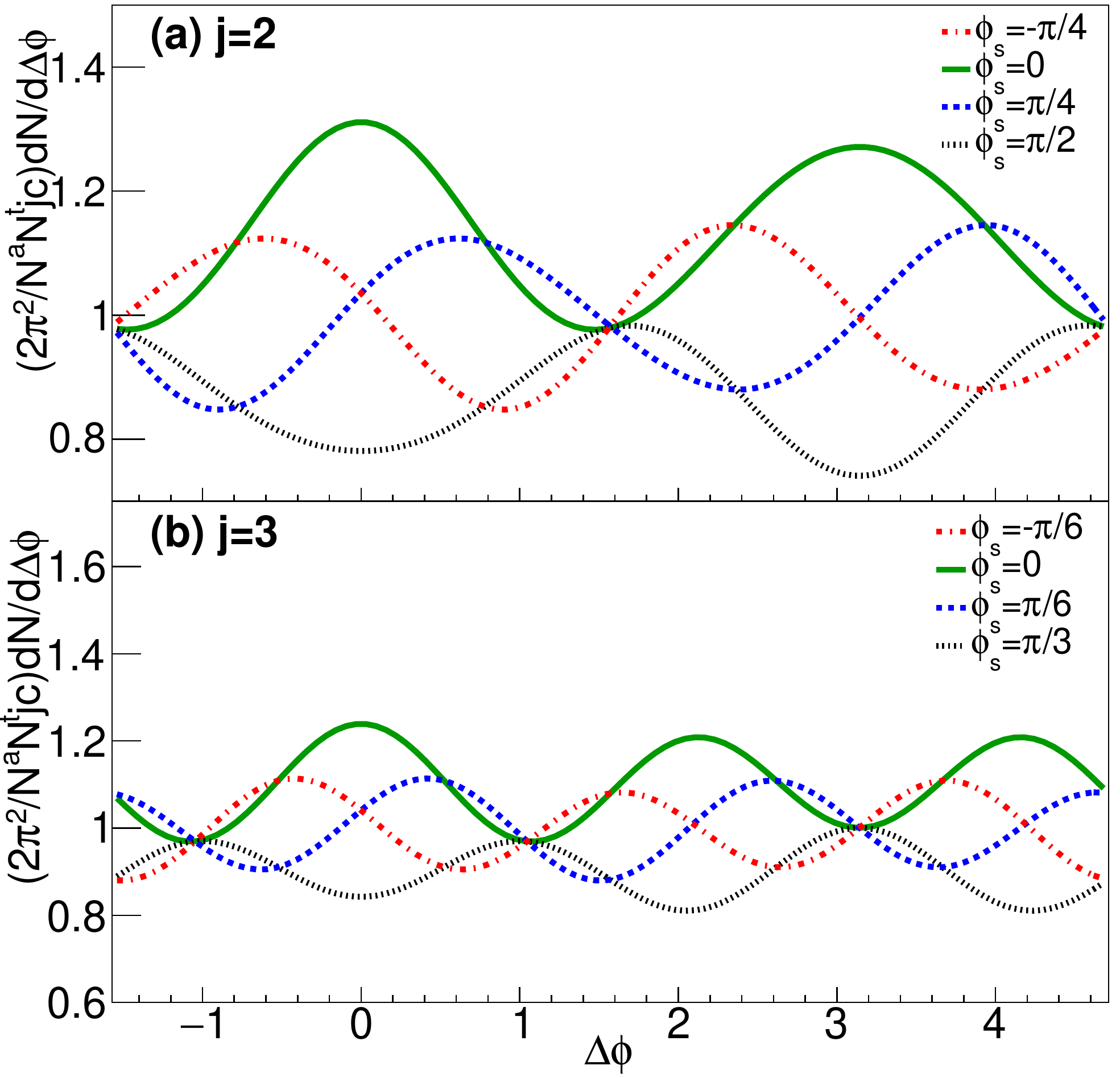}
}}
\caption{
Shape of correlations relative to the (a) $j=2$ participant plane with $c=\pi/6$ and (b) $j=3$ participant plane with $c=\pi/9$ for different orientations of the trigger relative to the participant plane for $v_2^{t}=v_2^{a}=v_3^{t}=v_3^{a}= 0.1$, $v_4^{t}=v_4^{a}=0.02$, $C_{4,0,2} = C_{4,2,2} = C_{2,4,2} = 0.1$, $R_{2,2}=0.8$, $R_{3,3}=0.6$, and $R_{4,2}=0.4$.  The $C_{n,m,j}$ mixing odd and even terms are assumed to be zero, as are the $C_{n,m,j}$ mixing odd terms of different orders.
}
\label{Fig:PhasShift}
\end{center}
\end{figure}

The assumptions used for deriving \eref{Eq:BFE} are that both the trigger and the associated particle are correlated with an event plane, which need not be the $j$th order participant plane, and that when averaged over events $\langle \sin(n (\psi_{j}^{reco} - \psi_{j}^{true})) \rangle = \langle \sin(n\psi_{n}+m\psi_{m}-(n+m)\psi_{j}) \rangle = 0$.  
Furthermore, we assume that the impact of event-by-event \vn fluctuations leading to correlations between \vn of different orders is negligible.
The degree of correlation between event planes of different orders is described by the $ C_{n,m,j}$.  This correlation need not arise from the same physical mechanism for the trigger and associated particles; it may be due to jet quenching for the trigger and flow for the associated particle. The $n$th order event plane resolution of the $j$th order event plane is shown by $R_{n,j}$. Note that the background shape in equation\ref{Eq:BFE} is different for different experiments even in the same collision energy, centrality, and \pT selections for the trigger and associated particles because $R_{n,j}$ depends on detector performance. 

The $ \tilde{w}_{n}^{t} = 0$ if $j\phi_s = n\pi$ where $n$ is an integer, such as the $\phi_s=0$ and $\phi_s=\pi/2$ cases investigated for $j=2$ in~\cite{Bielcikova:2003ku}.  The $ \tilde{w}_{n}^{t}$ are also zero if two regions with $\phi_s = \alpha$ and $\phi_s = -\alpha$ are summed, as in~\cite{Agakishiev:2014ada,Adare:2010mq,Nattrass:2016cln}.  \Fref{Fig:PhasShift} illustrates the effect of this phase shift.  This shift is crucial for understanding the background for triggers fixed in asymmetric regions relative to the event plane as in~\cite{Adare:2018wjb}, which could provide additional constraints for the path length dependence of energy loss.  It also may provide useful information for determining the shape of these correlations from a fit, such as in~\cite{Sharma:2015qra}.  
Note that the \vnt are only modified by \vnt with $n$ which are separated by multiples of $j$.
This is not due to partial correlation between participant planes of different orders but rather destructive interference of terms which are not multiples of $j$.  
For instance, for the second order event plane, $j=2$, \vnumeff{2} is modified by \vnum{2}, \vnum{4}, \vnum{6}... and \vnumeff{3} is modified by \vnum{1}, \vnum{3}, \vnum{5}...  In the latter case, the \vn are multiplied by the $ C_{n,m,j}$ and $R_{n,j}$, which are generally small except when $n$, $m$, and $j$ are even.  While the $R_{n,j}$ can be measured, most of the $ C_{n,m,j}$ are not generally known.  However, the formulation in \eqref{Eq:BFEP1} and \eqref{Eq:BFE} can be used to set limits on the higher order correlations because $ 0 < C_{n,m,j} < 1$.

\begin{figure*}
\begin{center}
\rotatebox{0}{\resizebox{17cm}{!}{
	\includegraphics{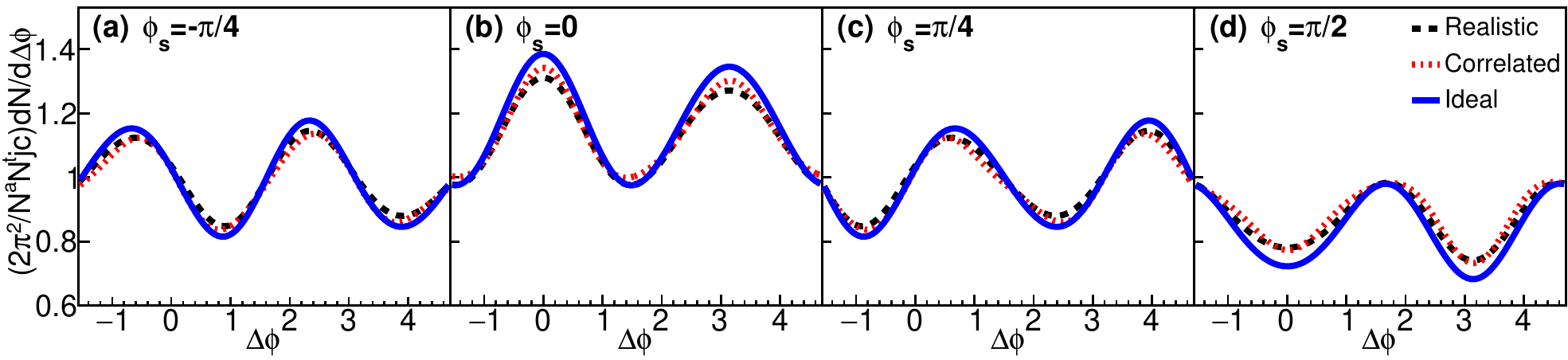}
}}
\caption{
Shape of correlations relative to the $j=2$ participant plane with $c=\pi/6$ for different orientations of the trigger relative to the participant plane for $v_2^{t}=v_2^{a}=v_3^{t}=v_3^{a}= 0.1$, $v_4^{t}=v_4^{a}=0.02$ comparing realistic reaction plane resolution and correlations between participant planes ($C_{4,0,2} = C_{4,2,2} = C_{2,4,2} = 0.1$, $R_{2,2}=0.8$, and $R_{4,2}=0.4$), ideal reaction plane resolution ($R_{2,2}=R_{4,2}=R_{6,4}=1$), and perfect correlation between even order participant planes ($C_{4,0,2} = C_{4,2,2} = C_{2,4,2} = 1$).
}
\label{Fig:Resolution}
\end{center}
\end{figure*}

\Fref{Fig:Resolution} shows the impact of realistic event plane resolution~\cite{Abelev:2014ffa} and possible correlations between event planes~\cite{Aad:2014fla} for the second order event plane.  For realistic correlations between the second and fourth order event planes, the impact of correlations is much smaller than the impact of the event plane resolution at each order.  Such terms may need to be taken into account, however, for precision measurements.  At higher order, cross terms such as $C_{6,4,2} = \langle \cos(6\psi_6 + 4\psi_4 - 10\psi_2 \rangle $ appear with a coefficient of $v_6$.  These terms may not be independently measured, but their impact can be estimated from a template fit to experimental data using \eref{Eq:BFE}.

The impact of \vnum{1} in such correlations is still unclear.  There are two contributions to the coefficient of $\cos(\Delta\phi)$, which is approximately
\begin{equation}
v_{1,1} =v_{1}^{flow,a} v_{1}^{flow,t} -  k\frac{p_T^a p_T^t}{N}\label{Eq:v1}
\end{equation}
where $v_{1}^{flow,a}$ and $v_{1}^{flow,t}$ are from rapidity-even hydrodynamical flow, k is a constant with respect to \dphi, $p_T^a$ is the momentum of the associated particle, $p_T^t$ is the momentum of the trigger particle, and $N$ is the event multiplicity.  There may also be a residual contribution from other non-flow effects
such as resonance decay, Bose-Einstein correlations, and jets.  The rapidity odd term is of particular interest to constrain the equation of state~\cite{Nara:2016phs}, but it is usually small at midrapidity.  Furthermore it has a sign change for pseudorapidity $\eta=0$ and is symmetric about $\eta=0$ for symmetric collisions, so its average is usually zero unless the measurement explicitly distinguishes between the directions of the incoming nuclei.
The fluctuations in initial nucleon position which lead to the other odd \vn also lead to a rapidity-even \vnum{1}~\cite{Teaney:2010vd}, although there are also contributions from the eccentricity in the initial state and nonlinear mixing between harmonics~\cite{Gardim:2014tya}.
Both rapidity-even flow and momentum conservation terms impact the background in \dhcs and it is unclear if they impact \jhcs.

The term $- k\frac{p_T^ap_T^t}{N}$  is from global momentum conservation, as derived in~\cite{Borghini:2000cm}.  This derivation assumed that momentum conservation is the only correlation in the collision.  The only contribution with this assumption is \vnum{1} because it is proportional to the dot product of the momenta, although there may be higher order corrections.  


While the $p_{T}$-integrated rapidity-even \vnum{1} due to flow times $p_{T}$, $\int v_1^{flow}(p_T)p_T dp_T$, is zero due to momentum conservation, it has been measured to be negative at low momenta and comparable to \vnum{2} and \vnum{3} at high momenta~\cite{Luzum:2010fb,ATLAS:2012at,Retinskaya:2012ky}.  This corresponds to a preferred direction in the collision, with high momentum particles preferentially in the opposite direction of low momentum particles.  The momentum conservation term was observed to be significant in these papers as well.  Both measurements use \dhcs with a large separation in pseudorapidity between trigger and associated particles and assume that non-flow contributions are negligible. To extract \vnum{1} as a function of momentum, $v_{1,1}$ is measured in several different momentum bins and fit to separate the momentum conservation and flow terms.  Note that \eref{Eq:v1} neglects event-by-event flow fluctuations.  The large separation in pseudorapidity suppresses contributions from hadrons from the same jet as the trigger hadron, however, there may be residual contributions from jets $\pi$ radians away from the trigger hadron in azimuth.  The measurement in~\cite{Luzum:2010fb} may still have residual contributions from hadrons in the same jet as the trigger hadron because the separation in pseudorapidity, $|\Delta\eta| = |\eta^a - \eta^t|<0.7$, is not wide enough to exclude all particles  since the width of the jet-like peak on the near side is  around 0.4~\cite{Agakishiev:2011st} at the lowest momenta.  The \vnum{1} measured may also be sensitive to the $\eta$ gap between the trigger and associated momenta.


In summary, the rapidity even \vnum{1} due to flow has been measured to be comparable to \vnum{2} and \vnum{3} and the global momentum conservation is also non-negligible, but there are not currently measurements which are reliable enough to subtract this contribution with precision in \dhcs.  Its subtraction in \jhcs is even more complicated, since only \vnum{2} has been measured for reconstructed jets.  We therefore urge caution with respect to the treatment of the rapidity-even \vnum{1} term.  The ZYAM method requires independent measurements of the \vn.  The reaction plane fit method described in~\cite{Sharma:2015qra} allows the inclusion of a \vnum{1} term and therefore could be used to reliably subtract this term.  It may also allow for more reliable measurements of this term, since contributions from jets are strongly suppressed.


\section{Conclusions}\label{Sec:Conclusions}
 \makeatletter{}We have derived the shape of the flow-modulated background in \dhjhcs changes when the trigger is fixed relative to an event plane at an arbitrary order $j$, including both the finite event plane resolution and realistic correlations between different order event planes.  There is a phase shift in this background when asymmetric regions about the event plane are studied.  The \vn in this form are only modified by the contributions from odd multiples of $j$, independent of the correlations between other order event planes.  For realistic correlations between event planes, we find only small effects from the correlation between participant planes of different orders.  We urge caution with respect to the treatment of the rapidity even \vnum{1} due to flow in such studies because this component is not constrained well by data. 
 \makeatletter{}
 
\section{Acknowledgements}
We are grateful to Jana Biel\v{c}ikova, Nicolas Borghini, Raymond Ehlers, 
ShinIchi Esumi, Matt Luzum, and Jaki Noronha-Hostler, and Jean-Yves Ollitrault for useful discussions and comments on the manuscript and to Roy Lacey, Anthony Timmins,  and Sergei Voloshin for useful discussions.  This work was supported in part by funding from the Division of Nuclear Physics of the U.S. Department of Energy under Grant No. DE-FG02-96ER40982. 

\bibliography{Bibliography,BibliographyNew}   
 \clearpage
\appendix*

 {
\onecolumngrid
\section{Derivations}\label{Sec:Derivations}
We follow the notation and terminology from~\cite{Bielcikova:2003ku}, expanding it for an arbitrary order participant plane and taking decorrelations between different order event planes into account.
The azimuthal anisotropy of single hadrons relative to 
the \jth order event plane is
\begin{equation} \label{Eq:distribution2}
 \frac{dN}{d(\phi-\psi_j)} = \frac{N}{2\pi} \Big (1+2 \sum_{n=1}^{\infty} v_{n} \cos(n(\phi-\psi_n)) \Big ) 
\end{equation}
\noindent where $N$ is the number of particles, $\phi$ is the position of the particle in azimuth, $\psi_n$ is the position of the $n$th order participant plane in azimuth, and 
$v_n = \langle \cos(n(\phi-\psi)) \rangle$ where $\psi_j$ need not equal $\psi_n$.  We assume that the azimuthal anisotropy of a jet can be similarly quantified and refer to the trigger particle or jet as a trigger in the following discussion.

To determine the azimuthal anisotropy between an associated particle and a trigger when the trigger azimuthal angle relative to the $j$th order event plane $\phi_s=\phi^{t} - \psi_j$ is restricted to $\phi_s-c < \phi_s < \phi_s+c$, we write equations like \eref{Eq:distribution2} for each, multiply them, integrate over possible angles between the reaction plane angle and the trigger position, and average over several events.  These integrals run from $\phi-\psi_j = \phi_s - c$ to $\phi-\psi_j = \phi_s + c$ for the \jth order event plane and there are $j$ integrals so the operator to integrate over this region is given by
\begin{equation}\label{Eq:integraloperator}
 \sum_{k=0}^{j-1} \int_{\phi_s - c + \frac{2\pi k}{j}}^{\phi_s +c + \frac{2\pi k}{j}} d(\phi-\psi_j).
\end{equation}
\noindent In the case where the measurement is done relative to the reconstructed participant plane, the operator in \eref{Eq:integraloperator} can be rewritten as
\begin{equation}\label{Eq:integraloperatorreco}
 \sum_{k=0}^{j-1} \int_{\phi_s - c + \psi_{j}^{reco}-\psi_{j}^{true}+ \frac{2\pi k}{j}}^{\phi_s +c + \psi_{j}^{reco}-\psi_{j}^{true} + \frac{2\pi k}{j}} d(\phi-\psi_j^{reco})
\end{equation}
by denoting the true participant plane $\psi_j = \psi_j^{true}$, writing $\phi-\psi_j^{reco}$ = $(\phi-\psi_j^{true})-(\psi_j^{true}-\psi_j^{reco})$ and changing the variable of integration.

For convenience, we define $x = \phi^t-\psi_j^{reco}$ where the superscript $t$ indicates that this is the position of the trigger, $\Delta\phi = \phi^a-\phi^t$, $\Delta\psi_{j}^{reco}= \psi_j^{reco}-\psi_j^{true}$ and $\Delta\psi_{ab}= \psi_a^{true}-\psi_b^{true}$.  We then write the distribution of trigger as
\begin{equation} \label{Eq:distributiontrig1}
 \frac{dN^t}{d(\phi^t-\psi_j)} = \frac{N^t}{2\pi} \Big (1+2 \sum_{n=1}^{\infty} v_{n}^{t} \cos(n(\phi^t-\psi_n)) \Big ) 
 = \frac{N^t}{2\pi} \Big (1+2 \sum_{n=1}^{\infty} v_{n}^{t} \cos(n(\phi^t-\psi_j+\psi_j-\psi_n)) \Big ),
\end{equation}
\noindent or
\begin{equation} \label{Eq:distributiontrig2}
 \frac{dN^t}{dx} = \frac{N^t}{2\pi} \Big (1+2 \sum_{n=1}^{\infty} v_{n}^{t} \cos(nx+n\Delta\psi_{jn}) \Big ).
\end{equation}
\noindent Similarly, the distribution of associated particles can be written 
\begin{equation}\label{Eq:distributionassoc1}
 \frac{dN^a}{d(\phi^a-\psi_j)} = \frac{N^a}{2\pi} \Big (1+2 \sum_{m=1}^{\infty} v_{m}^{a} \cos(m(\phi^a-\psi_m)) \Big ) 
 = \frac{N^a}{2\pi} \Big (1+2 \sum_{m=1}^{\infty} v_{m}^{a} \cos(m(\phi^t-\psi_j+\phi^a-\phi^t+\psi_j-\psi_m)) \Big ),
\end{equation}
\noindent or,
\begin{equation}\label{Eq:distributionassoc2}
 \frac{dN^a}{dx} = \frac{N^a}{2\pi} \Big (1+2 \sum_{m=1}^{\infty} v_{m}^{a} \cos(m(x+\Delta\phi+\Delta\psi_{jm})) \Big ).
\end{equation}

We then put these pieces together to get the background as a function of \dphi:
\begin{align}
 B(\Delta\phi) =& \frac{N^t N^a}{4\pi^2}  \sum_{k=0}^{j-1} \int_{\phi_s - c + \Delta\psi_{j}^{reco}+ \frac{2\pi k}{j}}^{\phi_s +c + \Delta\psi_{j}^{reco}+ \frac{2\pi k}{j}} dx \Big (1+2 \sum_{m=1}^{\infty} v_{m}^{a} \cos(mx+m\Delta\phi+m\Delta\psi_{jm}) \Big )\Big (1+2 \sum_{n=1}^{\infty} v_{n}^{t} \cos(nx+n\Delta\psi_{jn}) \Big ) \nonumber \\
 = & \frac{N^t N^a}{4\pi^2}  \sum_{k=0}^{j-1} \int_{\phi_s - c + \Delta\psi_{j}^{reco}+ \frac{2\pi k}{j}}^{\phi_s +c+ \Delta\psi_{j}^{reco} + \frac{2\pi k}{j}} dx 
 \Big( 1+2 \sum_{m=1}^{\infty} v_{m}^{a} \cos(mx+m\Delta\phi+m\Delta\psi_{jm}) +2 \sum_{n=1}^{\infty} v_{n}^{t} \cos(nx+n\Delta\psi_{jn}) \\
& + 4 \sum_{m=1}^{\infty} \sum_{n=1}^{\infty} v_{n}^{a} v_{m}^{t} \cos(mx+m\Delta\phi+m\Delta\psi_{jm})  \cos(nx+n\Delta\psi_{jn})  \nonumber
\Big).
\end{align}
\noindent We define the four terms as $b_1(\Delta\phi)$, $b_2(\Delta\phi)$, $b_3(\Delta\phi)$, and $b_4(\Delta\phi)$, respectively, as
\begin{align}
 b_1(\Delta\phi) =& \frac{N^t N^a}{4\pi^2}\sum_{k=0}^{j-1} \int_{\phi_s - c + \Delta\psi_{j}^{reco}+ \frac{2\pi k}{j}}^{\phi_s +c+ \Delta\psi_{j}^{reco} + \frac{2\pi k}{j}} dx \\
  b_2(\Delta\phi) = & \frac{N^t N^a}{2\pi^2}\sum_{k=0}^{j-1} \sum_{m=1}^{\infty} v_{m}^{a} \int_{\phi_s - c + \Delta\psi_{j}^{reco}+ \frac{2\pi k}{j}}^{\phi_s +c+ \Delta\psi_{j}^{reco} + \frac{2\pi k}{j}} dx  \cos(mx+m\Delta\phi+m\Delta\psi_{jm}) \nonumber \\
  b_3(\Delta\phi) =& \frac{N^t N^a}{2\pi^2}\sum_{k=0}^{j-1} \sum_{n=1}^{\infty} v_{n}^{t}  \int_{\phi_s - c + \Delta\psi_{j}^{reco}+ \frac{2\pi k}{j}}^{\phi_s +c+ \Delta\psi_{j}^{reco} + \frac{2\pi k}{j}} dx \cos(nx+n\Delta\psi_{jn})\nonumber  \\
  b_4(\Delta\phi) = & \frac{N^t N^a}{\pi^2}\sum_{k=0}^{j-1} \sum_{n=1}^{\infty}\sum_{m=1}^{\infty} v_{n}^{t} v_{m}^{a} \int_{\phi_s - c + \Delta\psi_{j}^{reco}+ \frac{2\pi k}{j}}^{\phi_s +c+ \Delta\psi_{j}^{reco} + \frac{2\pi k}{j}} dx  \cos(mx+m\Delta\phi+m\Delta\psi_{jm})  \cos(nx+n\Delta\psi_{jn}).
\end{align}
We consider each of them below.

\subsection{First term $b_1(\Delta\phi)$}
\begin{equation}
 b_1(\Delta\phi) = \frac{N^t N^a}{4\pi^2}\sum_{k=0}^{j-1} \int_{\phi_s - c + \Delta\psi_{j}^{reco}+ \frac{2\pi k}{j}}^{\phi_s +c+ \Delta\psi_{j}^{reco} + \frac{2\pi k}{j}} dx = \frac{N^t N^a}{2\pi^2}\sum_{k=0}^{j-1} c = \frac{N^t N^a j c}{2\pi^2}
\end{equation}

\subsection{Second term $b_2(\Delta\phi)$}
\begin{align}
 b_2(\Delta\phi) = & \frac{N^t N^a}{2\pi^2}\sum_{k=0}^{j-1} \sum_{m=1}^{\infty} v_{m}^{a} \int_{\phi_s - c + \Delta\psi_{j}^{reco}+ \frac{2\pi k}{j}}^{\phi_s +c+ \Delta\psi_{j}^{reco} + \frac{2\pi k}{j}} dx  \cos(mx+m\Delta\phi+m\Delta\psi_{jm}) \nonumber \\
  = & \frac{N^t N^a}{2\pi^2}\sum_{k=0}^{j-1} \sum_{m=1}^{\infty} \frac{v_{m}^{a}}{m} \sin(mx+m\Delta\phi+m\Delta\psi_{jm}) \Big|_{\phi_s - c + \Delta\psi_{j}^{reco}+ \frac{2\pi k}{j}}^{\phi_s +c+ \Delta\psi_{j}^{reco}+ \frac{2\pi k}{j}} \nonumber \\
  = & \frac{N^t N^a}{\pi^2}\sum_{k=0}^{j-1} \sum_{m=1}^{\infty} \frac{v_{m}^{a}}{m} \sin(mc)  \cos(m\phi_s + m\Delta\psi_{j}^{reco}+ \frac{2\pi km}{j}+m\Delta\phi+m\Delta\psi_{jm}) 
\end{align}
\noindent using
\begin{equation}
 \sin(a+b) - \sin(a-b) = 2 \cos(a)\sin(b).
\end{equation}
\noindent We can further simplify this using
\begin{equation}
 \cos(a+b+c) = \cos(a)\cos(b)\cos(c) - \cos(a)\sin(b)\sin(c)- \sin(a)\cos(b)\sin(c)- \sin(a)\sin(b)\cos(c).
\end{equation}
\noindent The fact that the average over events $\langle \Delta\psi_{j}^{reco} \rangle  = 0 $ and $\langle \Delta\psi_{jm} \rangle  = 0 $ and the fact that these distributions are symmetric about 0 means that $\langle\sin( m\Delta\psi_{j}^{reco}) \rangle  = 0 $ and $\langle \sin(m\Delta\psi_{jm}) \rangle  = 0 $ .  The $\Delta\psi_{j}^{reco} $ and $\Delta\psi_{jm}$ terms can then be pulled out:
\begin{align}
 b_2(\Delta\phi) = & \frac{N^t N^a}{\pi^2}\sum_{k=0}^{j-1} \sum_{m=1}^{\infty} \frac{v_{m}^{a}}{m} \sin(mc) \langle \cos(m\Delta\psi_{j}^{reco}) \rangle \langle \cos(m\Delta\psi_{jm}) \rangle \cos(m\phi_s + \frac{2\pi km}{j}+m\Delta\phi). 
\end{align}

We will investigate the term 
\begin{align}
 \sum_{k=0}^{j-1} \cos(m\phi_s + \frac{2\pi km}{j}+m\Delta\phi)
 = \sum_{k=0}^{j-1} \Big( \cos(m\phi_s+\frac{2\pi km}{j})\cos(m\Delta\phi) - \sin(m\phi_s+\frac{2\pi km}{j})\sin(m\Delta\phi)\Big).
\end{align}
We use the identity
\begin{equation}
  \sum_{k=0}^{j-1} \big( \cos(ma + \frac{2\pi km}{j}) +  i\sin(ma + \frac{2\pi km}{j}) \big) = e^{ima} \sum_{k=0}^{j-1} e^{\frac{2\pi km}{j}i} = \begin{cases}
                                                                j \cos(ma)+ij\sin(ma) & , m={\mathrm {multiple\ of\ }} j  \\
                                                                0 &, {\mathrm {otherwise}}.
                                                               \end{cases}
\end{equation}

We can then write
\begin{align}
 b_2(\Delta\phi) = & \frac{N^t N^a}{\pi^2} \sum_{m=1}^{\infty} \frac{v_{m}^{a}\delta_{m,mult\ j} j}{m} \sin(mc) \langle \cos(m\Delta\psi_{j}^{reco}) \rangle \langle \cos(m\Delta\psi_{jm}) \rangle \cos(m\phi_s + m\Delta\phi)
\end{align}
\noindent where $\delta_{m,mult\ j}$ indicates that $m$ is an integer $k$ times $j$.

We define the following variables to simplify the equations:
\begin{align}
 R_{n,j} &= \langle \cos(n \Delta\psi_{j}^{reco}) \rangle  = \langle \cos(n (\psi_{j}^{reco} - \psi_{j}^{true})) \rangle\\
 C(n,m,j) &= \langle \cos(n\psi_{n}+m\psi_{m}-(n+m)\psi_{j}) \rangle
\end{align}

We can then simplify and rearrange:
\begin{align}\label{Eq:term2}
 b_2(\Delta\phi) = & \frac{N^t N^a j}{\pi^2} \sum_{m=1}^{\infty} \frac{v_{m}^{a} \delta_{m,mult\ j} }{m} \sin(mc) R_{m,j} C_{m,0,j} \cos(m\phi_s + m\Delta\phi) \nonumber \\
 = & \frac{N^t N^a jc}{2\pi^2} \Bigg( 2
 \sum_{m=1}^{\infty} \frac{v_{m}^{a} \delta_{m,mult\ j} }{mc} \sin(mc)  R_{m,j} C_{m,0,j}  \big( \cos(m\phi_s) \cos(m\Delta\phi) - \sin(m\phi_s)\sin(m\Delta\phi) \big)
 \Bigg) \nonumber \\
 = & \frac{N^t N^a jc}{2\pi^2} \Bigg( 2
 \sum_{k=1}^{\infty} \frac{v_{jk}^{a}  }{jkc} \sin(jkc)  R_{jk,j} C_{jk,0,j}  \big( \cos(jk\phi_s) \cos(jk\Delta\phi) - \sin(jk\phi_s)\sin(jk\Delta\phi) \big)
 \Bigg)
\end{align}

\subsection{Third term $b_3(\Delta\phi)$}
\begin{align}
 b_3(\Delta\phi) &= \frac{N^t N^a}{2\pi^2}\sum_{k=0}^{j-1} \sum_{n=1}^{\infty} v_{n}^{t}  \int_{\phi_s - c + \Delta\psi_{j}^{reco}+ \frac{2\pi k}{j}}^{\phi_s +c+ \Delta\psi_{j}^{reco} + \frac{2\pi k}{j}} dx \cos(nx+n\Delta\psi_{jn})\nonumber \\
  &= \frac{N^t N^a}{2\pi^2}\sum_{k=0}^{j-1} \sum_{n=1}^{\infty} \frac{v_{n}^{t}}{n}   \sin(nx+n\Delta\psi_{jn}) \Big|_{\phi_s - c + \Delta\psi_{j}^{reco}+ \frac{2\pi k}{j}}^{\phi_s +c+ \Delta\psi_{j}^{reco}+ \frac{2\pi k}{j}} \nonumber\\
  &= \frac{N^t N^a}{\pi^2}\sum_{k=0}^{j-1} \sum_{n=1}^{\infty} \frac{v_{n}^{t}}{n} \sin(nc)  \cos(n\phi_s  + n\Delta\psi_{j}^{reco}+ \frac{2\pi k n}{j}+n\Delta\psi_{jn})\nonumber\\
  &= \frac{N^t N^a}{\pi^2}\sum_{k=0}^{j-1} \sum_{n=1}^{\infty} \frac{v_{n}^{t}}{n} \sin(nc)  \langle\cos(n\Delta\psi_{j}^{reco})\rangle \langle\cos(n\Delta\psi_{jn})\rangle \cos(n\phi_s  + \frac{2\pi k n}{j})\nonumber\\
  &= \frac{N^t N^a}{\pi^2} \sum_{n=1}^{\infty} \frac{v_{n}^{t}\delta_{n,mult\ j} j }{n} \sin(nc)  \langle\cos(n\Delta\psi_{j}^{reco})\rangle \langle\cos(n\Delta\psi_{jn})\rangle \cos(n\phi_s)
\end{align}
\noindent following the same logic as for the second term.  Again we simplify and rearrange, including a shift of indices
\begin{equation}
 b_3(\Delta\phi) = \frac{N^t N^ajc}{2\pi^2} 2 \sum_{k=1}^{\infty}  \frac{v_{jk}^{t}  }{jkc} \sin(nc) R_{jk,j} C_{jk,0,j} \cos(jk\phi_s)
\end{equation}

\subsection{Fourth term $b_4(\Delta\phi)$}
\begin{equation}
 b_4(\Delta\phi) = \frac{N^t N^a}{\pi^2}\sum_{k=0}^{j-1} \sum_{n=1}^{\infty}\sum_{m=1}^{\infty} v_{n}^{t} v_{m}^{a} \int_{\phi_s - c + \Delta\psi_{j}^{reco}+ \frac{2\pi k}{j}}^{\phi_s +c+ \Delta\psi_{j}^{reco} + \frac{2\pi k}{j}} dx  \cos(mx+m\Delta\phi+m\Delta\psi_{jm})  \cos(nx+n\Delta\psi_{jn})
 \end{equation}
We consider $n=m$ and $n \ne m$ terms separately.

\subsubsection{$n=m$}
 We use the integral
\begin{equation}
 \int \cos(n(x+a))\cos(n(x+b)) dx =\frac{x}{2} \cos(n(a-b)) +  \frac{1}{4} \frac{\sin(n(a+b+2x))}{n} + C\label{Eq:Integral4}
\end{equation} 
\noindent to simplify
\begin{align}
 b_4(\Delta\phi)& = \frac{N^t N^a}{\pi^2}\sum_{k=0}^{j-1} \sum_{n=1}^{\infty} v_{n}^{t} v_{n}^{a}  \int_{\phi_s - c + \Delta\psi_{j}^{reco}+ \frac{2\pi k}{j}}^{\phi_s +c+ \Delta\psi_{j}^{reco} + \frac{2\pi k}{j}} dx  \cos(nx+n\Delta\phi+n\Delta\psi_{jn})  \cos(nx+n\Delta\psi_{jn})\nonumber \\
 &= \frac{N^t N^a}{\pi^2}\sum_{k=0}^{j-1} \sum_{n=1}^{\infty} v_{n}^{t} v_{n}^{a} \Big( c \cos(n\Delta\phi) +   \frac{\sin(2nc)\cos(2n\Delta\psi_{jn}+2n\phi_s + 2n \Delta\psi_{j}^{reco}+ 2n\frac{2\pi k}{j}+n\Delta\phi)}{2n}  \Big) \nonumber\\
 &= \frac{N^t N^a}{\pi^2}\sum_{k=0}^{j-1} \sum_{n=1}^{\infty} v_{n}^{t} v_{n}^{a} \Big( c \cos(n\Delta\phi) +   \frac{\sin(2nc)\langle \cos(2n\Delta\psi_{jn}) \rangle \langle \cos(2n \Delta\psi_{j}^{reco}) \rangle \cos(2n\phi_s + 2n\frac{2\pi k}{j}+n\Delta\phi)}{2n}  \Big)\nonumber \\
 &= \frac{N^t N^aj}{\pi^2} \sum_{n=1}^{\infty} v_{n}^{t} v_{n}^{a} \Big(  c \cos(n\Delta\phi) +   \frac{\delta_{2n,mult\ j}  \sin(2nc)\langle \cos(2n\Delta\psi_{jn}) \rangle \langle \cos(2n \Delta\psi_{j}^{reco}) \rangle \cos(2n\phi_s+n\Delta\phi)}{2n}  \Big) \nonumber\\
 &= \frac{N^t N^ajc}{2\pi^2} 2\sum_{n=1}^{\infty} v_{n}^{t} v_{n}^{a} \Big(   \cos(n\Delta\phi) +   \frac{\delta_{2n,mult\ j}  \sin(2nc) C_{n,n,j} R_{2n,j} \cos(2n\phi_s+n\Delta\phi)}{2nc}  \Big). \label{Eq:term4neqm}
 \end{align}

 \subsubsection{$n\ne m$}
 We use the integral
\begin{align}
 \int \cos(n(x+a))\cos(m(x+b)) dx  = &\frac{1}{2} \frac{\sin((m-n)x+na-mb)}{n-m} +  \frac{1}{2} \frac{\sin((m+n)x+na+mb)}{n+m} + C \nonumber \\
 \int_{\alpha-\beta}^{\alpha+\beta} \cos(n(x+a))\cos(m(x+b)) dx  = &\frac{1}{2} \frac{\sin((m-n)\beta)\cos((m-n)\alpha+na-mb)}{n-m} \nonumber \\
& +  \frac{1}{2} \frac{
 \sin((m+n)\beta)
 \cos((m+n)\alpha+na+mb)
 }{n+m} 
\end{align} 
\noindent to simplify
\begin{align}
 b_4(\Delta\phi) =& \frac{N^t N^a}{\pi^2}\sum_{k=0}^{j-1} \sum_{n=1}^{\infty}\sum_{m=1}^{\infty} v_{n}^{t} v_{m}^{a} \int_{\phi_s - c + \Delta\psi_{j}^{reco}+ \frac{2\pi k}{j}}^{\phi_s +c+ \Delta\psi_{j}^{reco} + \frac{2\pi k}{j}} dx  \cos(mx+m\Delta\phi+m\Delta\psi_{jm})  \cos(nx+n\Delta\psi_{jn}) \nonumber\\
  =& \frac{N^t N^a}{\pi^2}\sum_{k=0}^{j-1} \sum_{n=1}^{\infty}\sum_{m=1}^{\infty} v_{n}^{t} v_{m}^{a}
\Big(
\frac{\sin((n-m)c)\cos((n-m)(\phi_s + \Delta\psi_{j}^{reco}+ \frac{2\pi k}{j}) -n\Delta\psi_{jn} +m\Delta\psi_{jm} + m\Delta\phi)}{n-m}\nonumber \\
&+  \frac{\sin((m+n)c)\cos((n+m)(\phi_s + \Delta\psi_{j}^{reco}+ \frac{2\pi k}{j}  ) +n\Delta\psi_{jn} +m\Delta\psi_{jm}+m\Delta\phi)}{n+m}
 \Big)\nonumber \\
 =&\frac{N^t N^a j c}{2 \pi^2} 2 \sum_{n=1}^{\infty}\sum_{m=1}^{\infty} v_{n}^{t} v_{m}^{a} 
 \Big(
\frac{\delta_{n-m,mult j}\sin((n-m)c)\cos((n-m)\phi_s- m\Delta\phi)R_{n-m,j}C_{n,-m,j}}{(n-m)c}\nonumber\\
&+ \frac{\delta_{n+m,mult j}\sin((m+n)c)\cos((n+m)\phi_s+ m\Delta\phi)R_{n+m,j}C_{n,m,j}}{(n+m)c}
 \Big).\label{Eq:term4nn}
 \end{align}
 
\subsubsection{$n=m$ and $n \ne m$ combined}
 Note that the second term in \eref{Eq:term4neqm} gets folded in to the $n+m$ term in \eref{Eq:term4nn}.  We add the term and shift indices:
 \begin{align}
  b_4(\Delta\phi)& =\frac{N^t N^ajc}{2\pi^2} 2\sum_{n=1}^{\infty} v_{n}^{a} \Bigg(   v_{n}^{t}\cos(n\Delta\phi) 
  + \sum_{k=1}^{\infty} (v_{jk+n}^{t}C_{|jk+n|,n,j}+v_{|jk-n|}^{t}C_{|jk-n|,n,j}) \frac{\sin(jkc)\cos(jk\phi_s- n\Delta\phi) R_{k,j}}{kc}
  \Bigg)
 \end{align}

\subsection{Putting it all together}
We want to write our equation in the form
\begin{equation}
 B(\Delta\phi) = \tilde{B}\Bigg(1+
 2\sum_{n=1}^{\infty} v_{n}^{a} \big( \tilde{v}_{n}^{t} \cos(n \Delta\phi) + \tilde{w}_{n}^{t} \sin(n \Delta\phi)\big)
\Bigg).
\end{equation}
\noindent By evaluating the previous terms and comparing the sections with $\cos(n \Delta\phi)$ and $\sin(n \Delta\phi)$ dependence, we can see
\begin{align}
 \tilde{B} &= \frac{N^t N^a j c}{2\pi^2} \Big(1
 +  2 \sum_{k=1}^{\infty}  \frac{v_{jk}^{t}  }{jkc} \sin(jkc) R_{jk,j} C_{jk,0,j} \cos(jk\phi_s)
 \Big),\nonumber \\
 \tilde{v}_{n}^{t} &= 
 \frac{
 v_{n} + \frac{\delta_{n,mult\ j} }{nc} \sin(nc) R_{n,j} C_{n,0,j}   \cos(n\phi_s)+ \sum_{k=1}^{\infty} (v_{jk+n}^{t}C_{|jk+n|,n,j}+v_{|jk-n|}^{t}C_{|jk-n|,n,j}) \frac{\sin(jkc)\cos(jk\phi_s) R_{jk,j}}{jkc}}{
 1 +  2 \sum_{k=1}^{\infty}  \frac{v_{jk}^{t}  }{jkc} \sin(nc) R_{jk,j} C_{jk,0,j} \cos(jk\phi_s)
 }\\
 \tilde{w}_{n}^{t} &= \frac{
 \frac{\delta_{n,mult\ j} }{nc} \sin(nc) R_{n,j} C_{n,0,j}   \sin(n\phi_s)+ \sum_{k=1}^{\infty}  (v_{jk+n}^{t}C_{|jk+n|,n,j}+v_{|jk-n|}^{t}C_{|jk-n|,n,j}) \frac{\sin(jkc)\sin(jk\phi_s) R_{jk,j}}{jkc}}{
 1 +  2 \sum_{k=1}^{\infty}  \frac{v_{jk}^{t}  }{jkc} \sin(nc) R_{jk,j} C_{jk,0,j} \cos(jk\phi_s)
 }\nonumber
\end{align}

}

\end{document}